\documentclass{PoS}
\pdfoutput=1
\def\apj{ApJ}%
\def\aap{A\&A}%
\def\jcap{JCAP}%
\def\mnras{MNRAS}%
\def\prd{Phys.\ Rev.\ D}%
\def\prl{Phys.\ Rev.\ Lett.}%
\def\physrep{Phys.\ Rep.}%
\def\repprogphys{Rep.\ Prog.\ Phys.}%
\title{Dark stars: structure, evolution and impacts upon the high-redshift Universe}

\ShortTitle{Dark stars: structure, evolution and impacts upon the high-redshift Universe}

\author{Pat Scott\\
        Department of Physics, McGill University,\\
        3600 rue University, Montr\'eal, QC, H3A 2T8, Canada\\
        E-mail: \email{patscott@physics.mcgill.ca}}

\abstract{The most compelling and popular models for dark matter predict that it should congregate and annihilate in stellar cores.  Stars where annihilation contributes substantially to the total energy budget look very different to those with which we are familiar.  Here I explain the general features of stars modified by dark matter annihilation with the help of a series of grids of `dark' stellar evolutionary models, and describe the public code with which they were computed.  I go on to discuss possible impacts of dark stars on the high-redshift Universe, including the history of reionisation.  The preliminary reionisation calculations reproduced here are based on dedicated models for dark star atmospheres, and for the stellar populations to which dark stars would belong.}

\FullConference{Cosmic Radiation Fields: Sources in the early Universe\\
		 November 9-12, 2010\\
		 DESY, Germany}

\begin{document}

\section{Introduction}

Weakly-interacting massive particles (WIMPs) are amongst the most promising candidates for dark matter \cite{Jungman96, Bergstrom00, Bertone05}.  In the standard scenario, WIMPs are Majorana particles that self-annihilate with a cross-section of $\sim3\times10^{-26}$\,cm$^3$\,s$^{-1}$, leading to a thermal relic dark matter abundance very similar to the observed value.  If such WIMPs are somehow confined within stellar cores, the resultant annihilation can have a substantial effect on stellar properties \cite{SalatiSilk,BouquetSalati}.  The idea that such `dark stars' could exist has been pursued rather vigorously in recent years, with attention given to white dwarfs \cite{MoskalenkoWai,Bertone08,McCullough10,Hooper10}, neutron stars \cite{Bertone08,Fairbairn10,Kouvaris10} and main-sequence stars \cite{Scott07,Fairbairn08,Scott09,Casanellas09,Scott09proc} in our own Galaxy, as well as to the first stars (Pop III) \cite{Spolyar08,Iocco08a,Freese08a,Iocco08b,Freese08b,Yoon08,Taoso08,Freese09,Natarajan09,Ripamonti09,Spolyar09,Iocco09,Umeda10,Sivertsson10,Ripamonti10,Gondolo10}.  The effect of dark matter annihilation on Pop III stars has been examined in the context of broader astronomical observations, including reionisation \cite{Schleicher, Venk11}, HST/JWST deep-field magnitudes and galaxy spectra \cite{Zackrisson10a,FreeseSMDS,Zackrisson10b}.

For the purposes of these proceedings, and more generally, I define a dark star as `any stellar object whose structure or evolution has been affected by dark matter annihilation'.  Other authors prefer to use the term `dark star' to refer exclusively to the dark-matter dominated phase that may occur during the formation of Pop III stars (e.g.~\cite{Spolyar08}).  The distinction is essentially just semantic, but important for readers' understanding of the literature, should they choose to delve into it.  These proceedings are not intended as a comprehensive review of dark stars as a field of active research, rather as simply a set of summarising notes of the (semi-review) presentation given at CRF2010.

\begin{figure}
\centering
\includegraphics[trim = 0 0 305 630, clip=true]{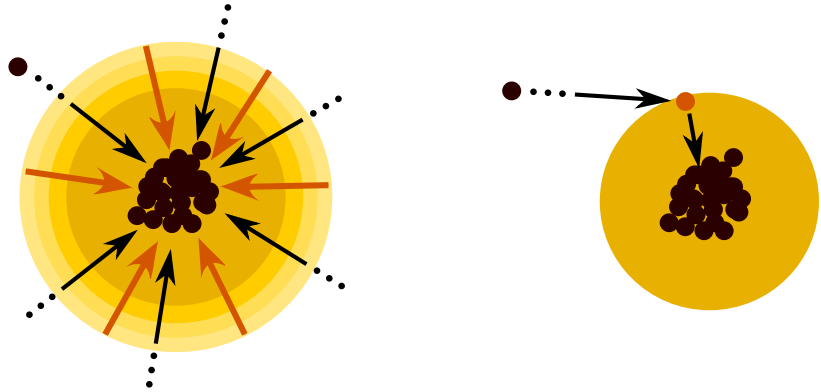}
\caption{A schematic representation of the two physical mechanisms by which dark stars acquire dark matter.  Gravitational contraction (left) occurs where baryonic collapse of a star-forming gas cloud steepens the gravitational potential, drawing dark matter in along with the baryons.  Nuclear scattering (right) relies on weak scattering events between WIMPs and nucleons, whereby dark matter particles lose energy and become gravitationally bound to the star.}
\label{fig1}
\end{figure}

Dark matter can find its way into stellar cores via two distinct physical mechanisms, as outlined in Fig.~\ref{fig1}.  The first is by the gravitational contraction of a baryonic gas cloud, as it cools and collapses during star formation.  The baryonic contraction steepens the gravitational potential, drawing dark matter into the centre of collapsing cloud.  The contraction of the dark matter need not be entirely adiabatic for this to occur, although adiabaticity is typically assumed.  This mechanism can only occur during formation of a star, so the resulting population of dark matter cannot be replaced by the same mechanism after it annihilates away.  The second mechanism is by nuclear scattering, whereby WIMPs scatter weakly off nucleons in the star, lose kinetic energy and become gravitationally bound to the star.  They then follow bound orbits to return and scatter repeatedly, losing more energy and settling down to the stellar core.  This mechanism can continue to be efficient for as long as an appropriate population of dark matter exists for the star to capture from in the halo surrounding it; in this way, capture via nuclear scattering may in principle replenish a star's population of dark matter indefinitely.  

\begin{figure}[p]
\begin{minipage}[t]{0.48\linewidth}
\centering
\includegraphics[width=0.96\linewidth, trim = 0 0 0 30, clip=true]{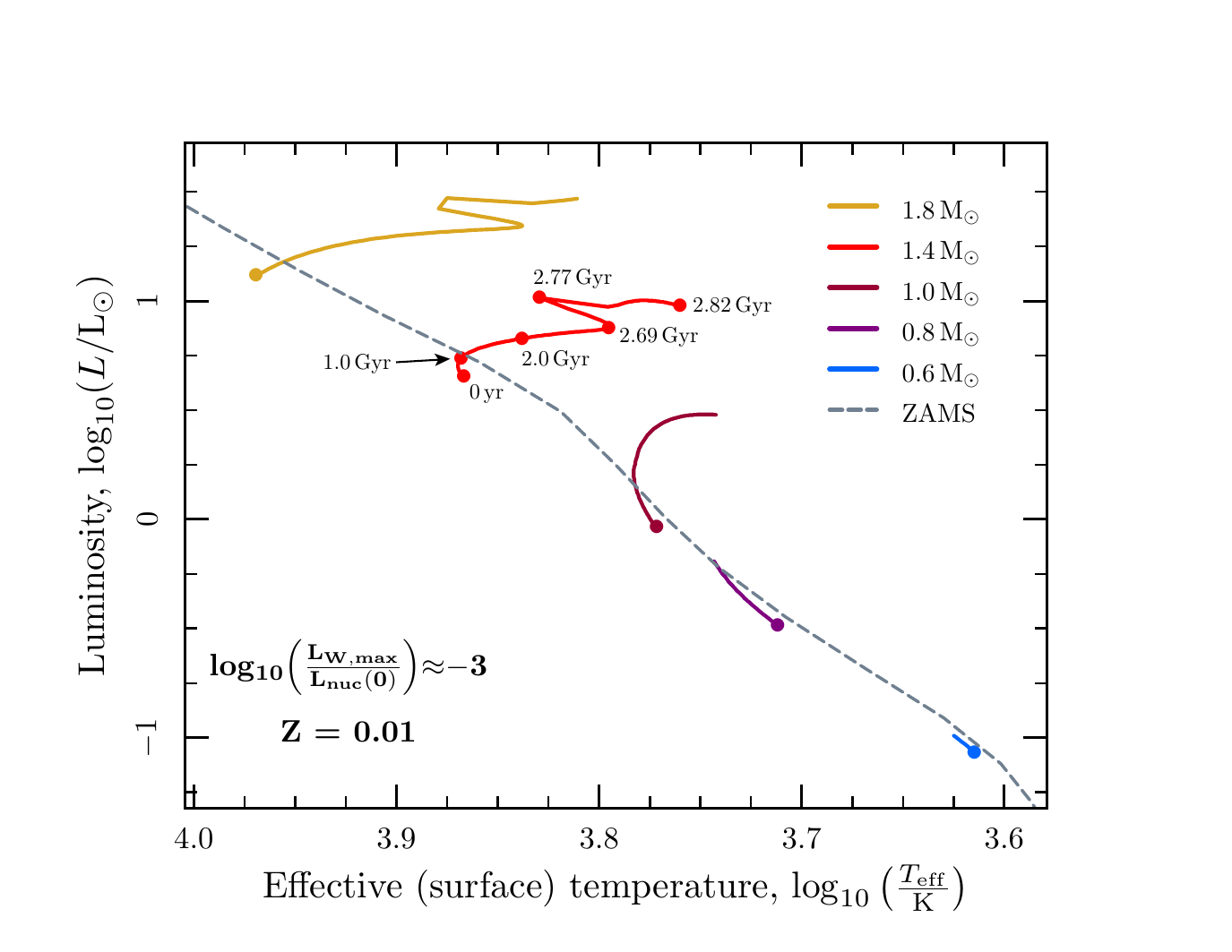}
\includegraphics[width=0.96\linewidth, trim = 0 0 0 30, clip=true]{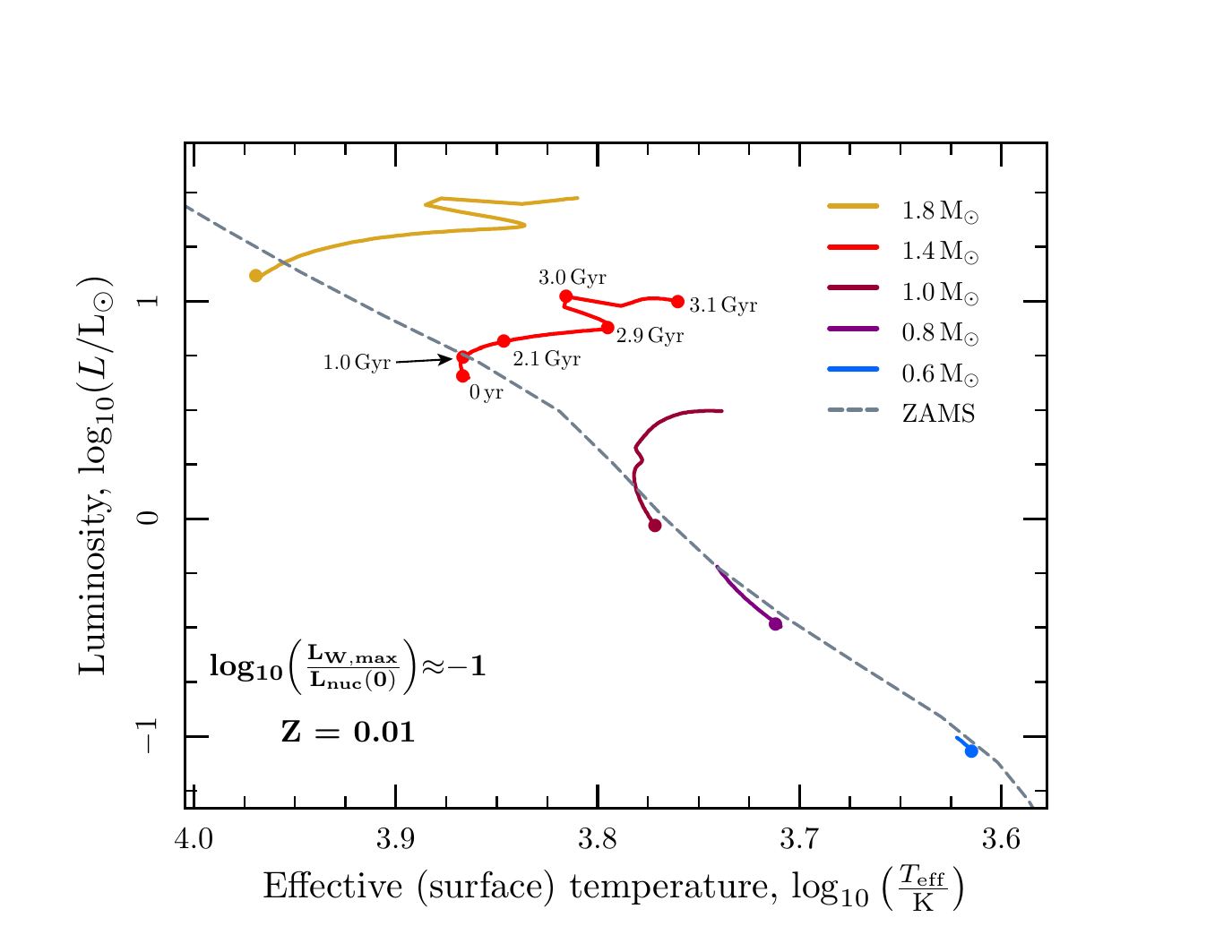}
\includegraphics[width=0.96\linewidth, trim = 0 0 0 30, clip=true]{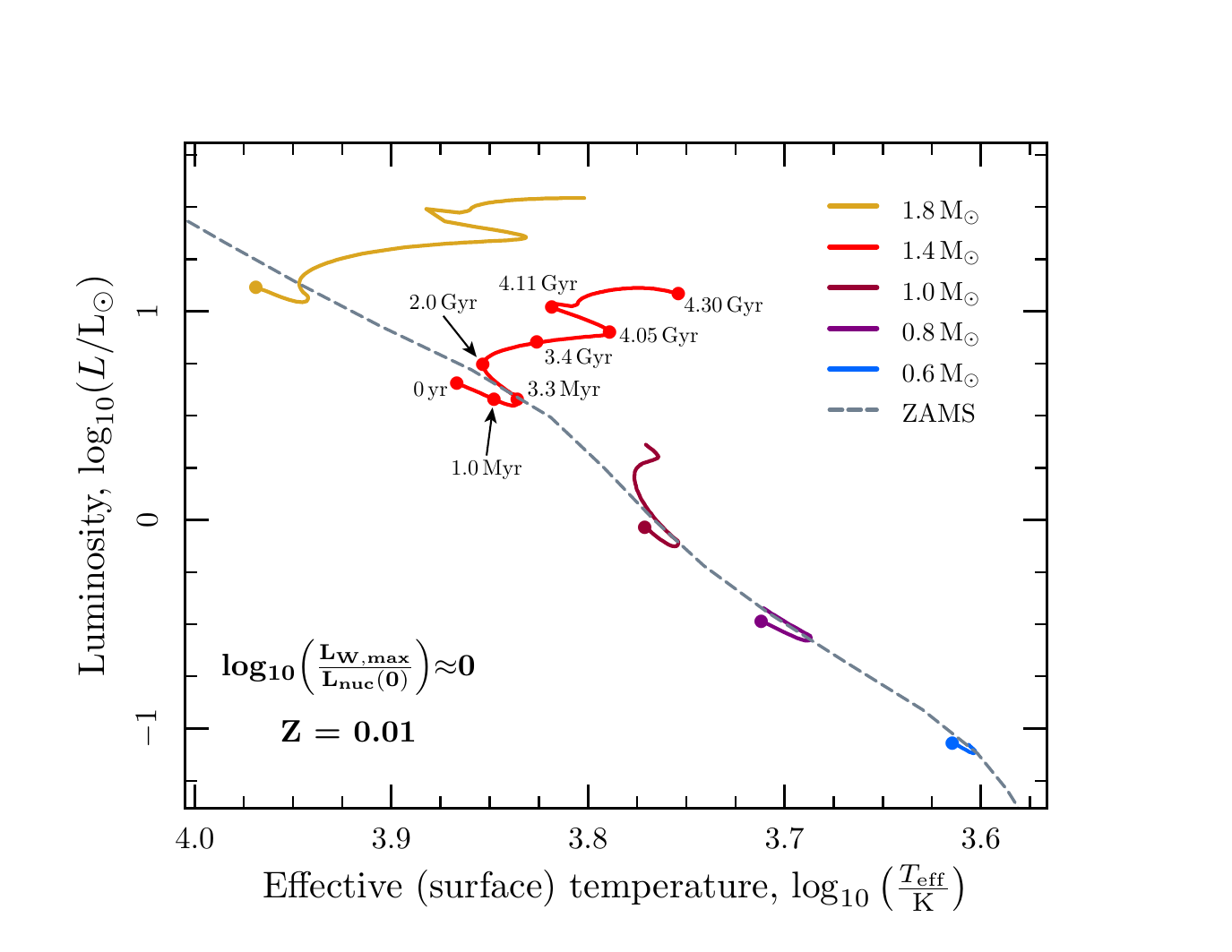}
\includegraphics[width=0.96\linewidth, trim = 0 0 0 30, clip=true]{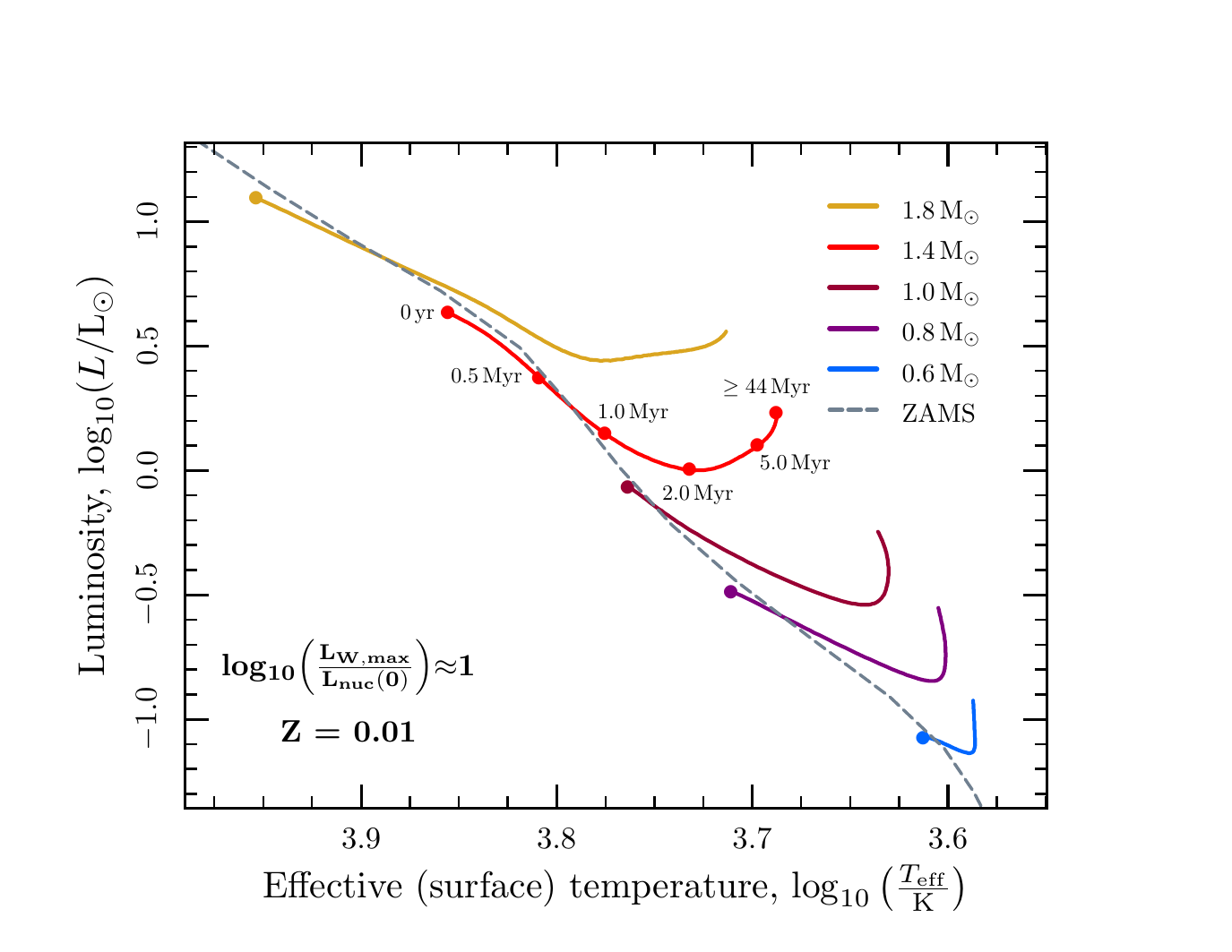}
\end{minipage}
\hspace{0.04\linewidth}
\begin{minipage}[t]{0.48\linewidth}
\centering
\includegraphics[width=0.96\linewidth, trim = 0 0 0 30, clip=true]{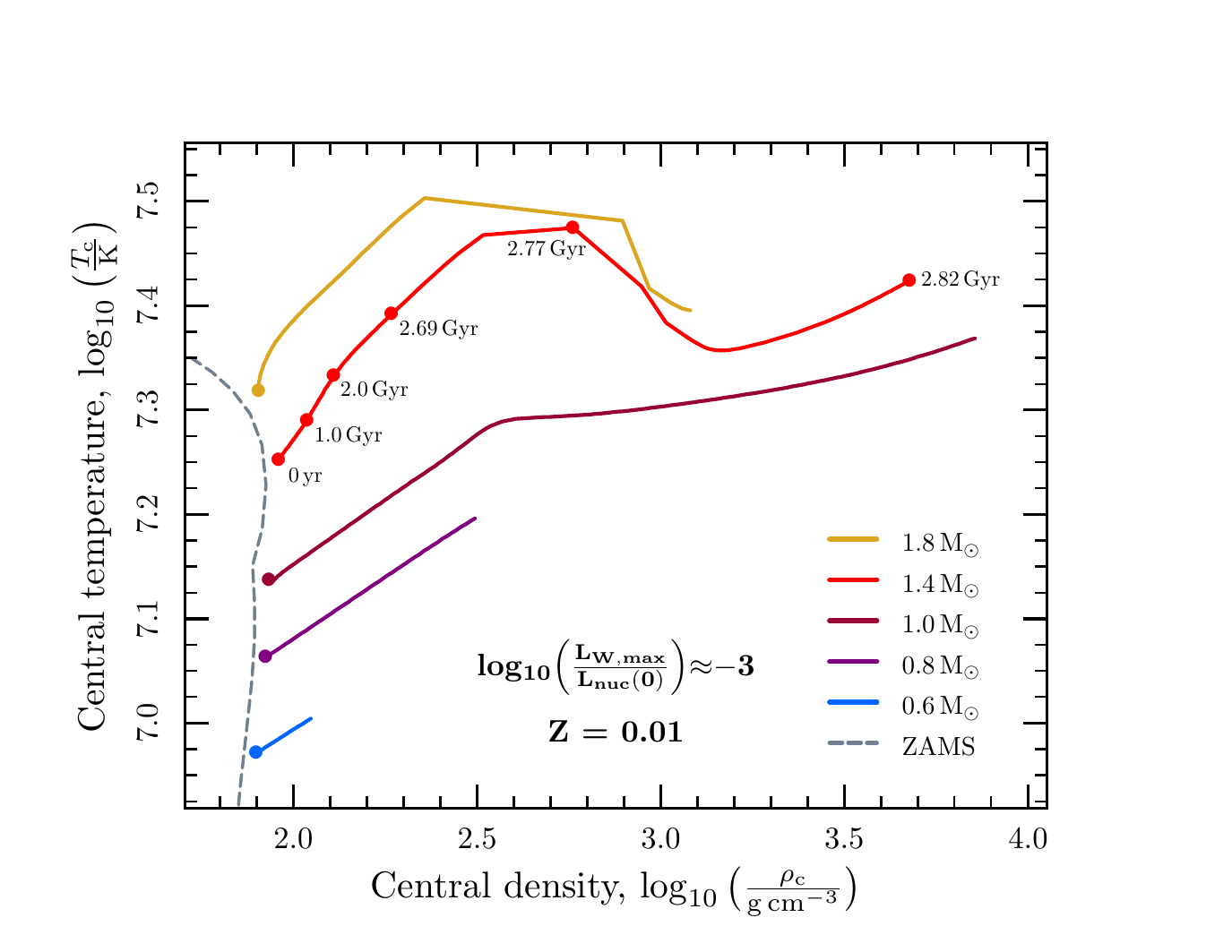}
\includegraphics[width=0.96\linewidth, trim = 0 0 0 30, clip=true]{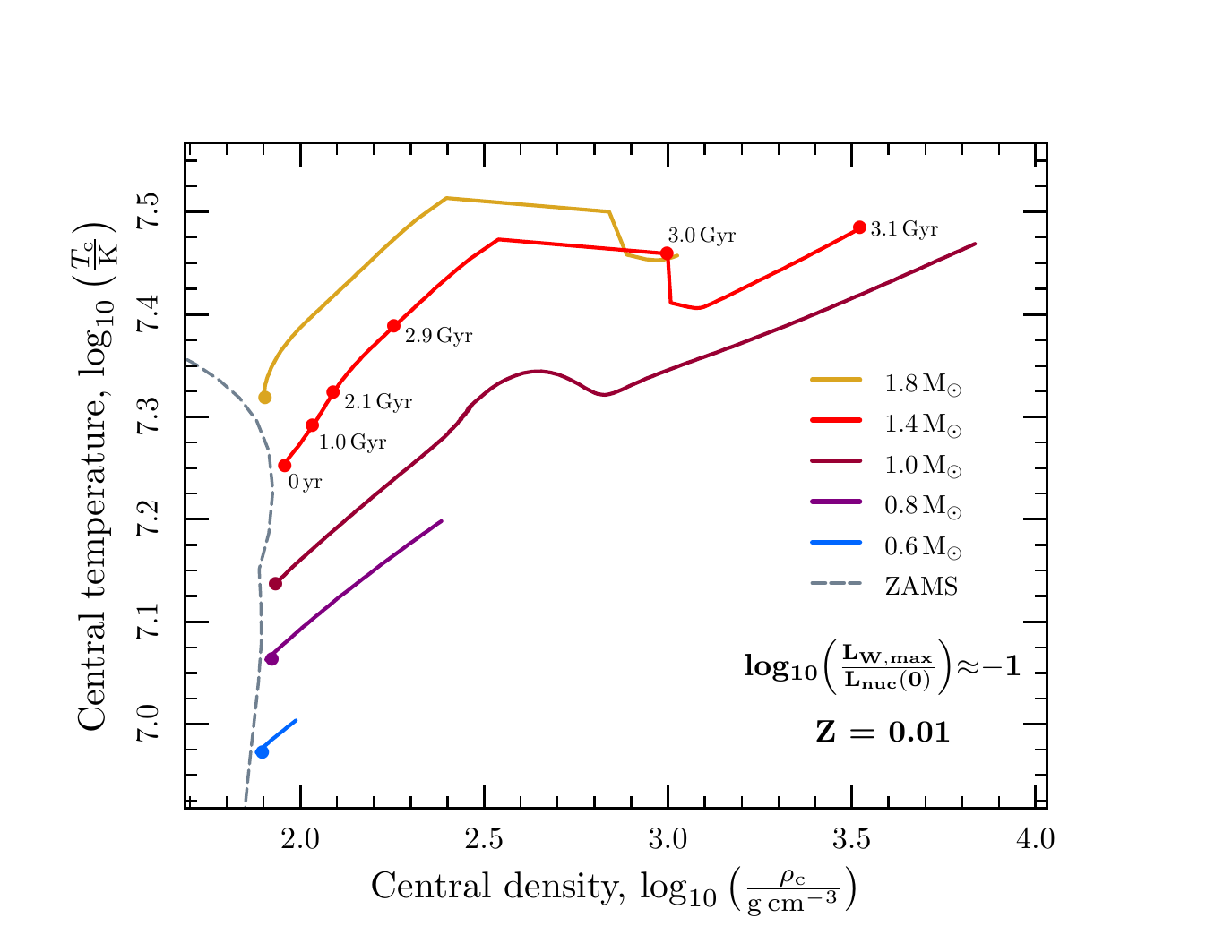}
\includegraphics[width=0.96\linewidth, trim = 0 0 0 30, clip=true]{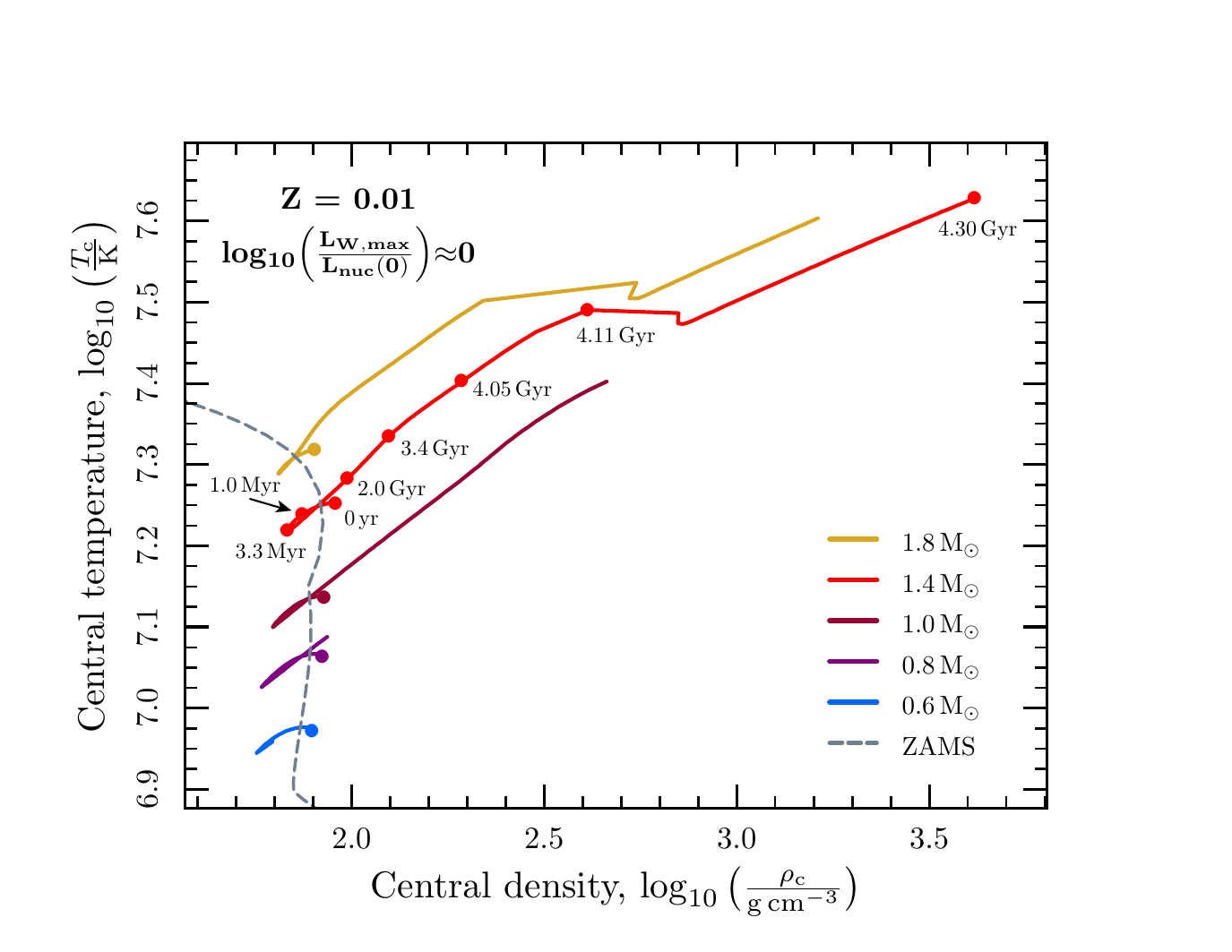}
\includegraphics[width=0.96\linewidth, trim = 0 0 0 30, clip=true]{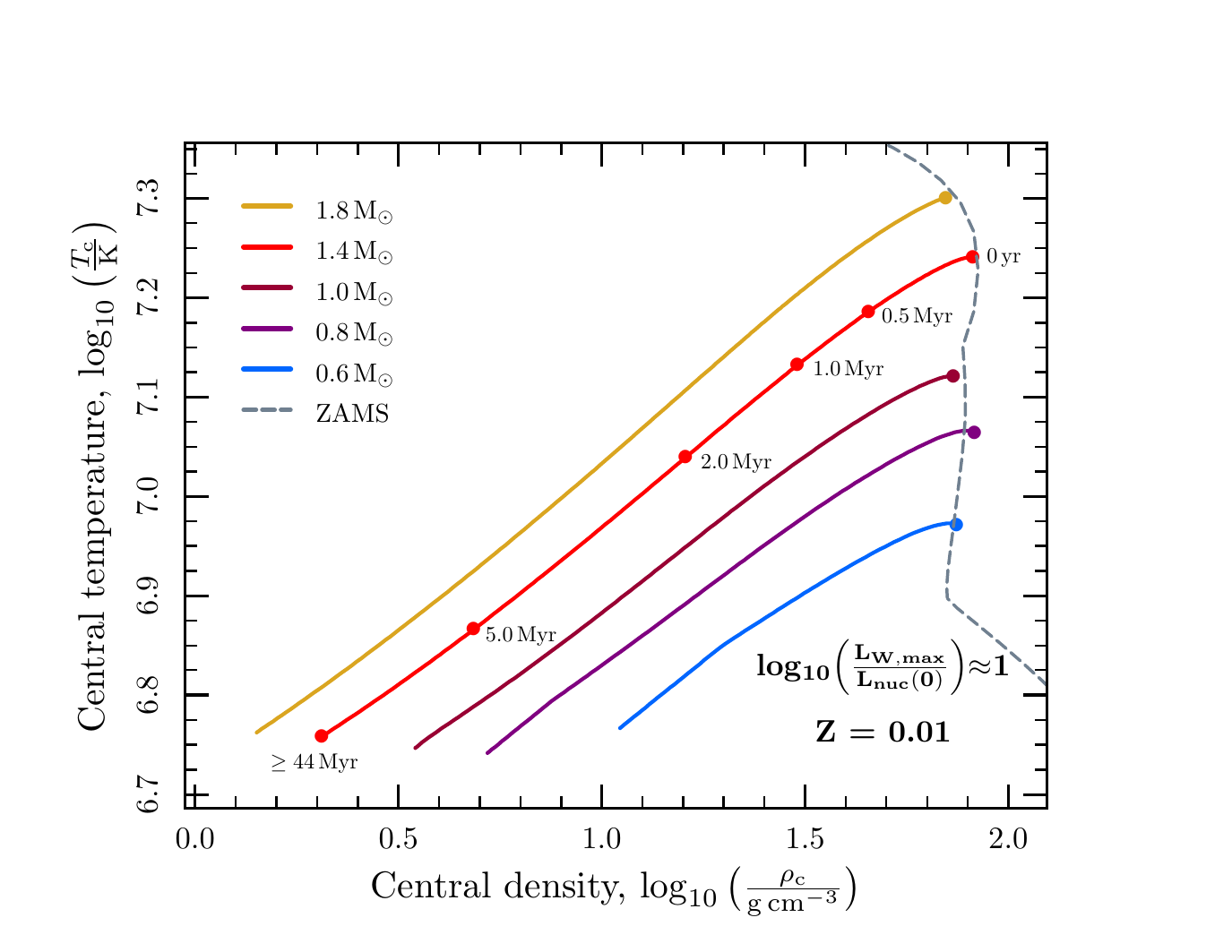}
\end{minipage}
\caption{Example evolutionary tracks (left) and central equation-of-state diagrams (right) for a grid of low-mass Pop II stars, when evolved from the main sequence with different amounts of dark matter capture by nuclear scattering.  From top to bottom, models exhibit sufficient capture to provide $10^{-3}$, $10^{-1}$, $1$ and $10$ times as much energy from WIMP annihilation as from nuclear fusion.  For details, see Ref.~\cite{Scott09}.  Models were computed using the \textsf{DarkStars} code \cite{Scott09proc}.  Qualitatively similar changes occur for higher masses and other metallicities.  Stars that are strongly affected by dark matter annihilation \emph{before} reaching the main sequence approach similar Hayashi-like solutions, but descend from higher on the Hayashi track instead of ascending to them from the main sequence.}
\label{fig2}
\end{figure}

\section{Structure and evolution of dark stars}

Regardless of the physical mechanism bringing dark matter into a star, the total amount present, and therefore the rate at which it injects energy into the centre of the star, are what ultimately determine the impact upon stellar structure and evolution.  Around 90\% of the rest mass of each WIMP is injected into the gas locally as additional heat \cite{Scott09}, so WIMP annihilation acts as an additional luminosity alongside nuclear burning.  This additional term is implemented in the open source dark stellar evolution code \textsf{DarkStars} \cite{Scott09proc}, along with highly detailed treatments of the capture of WIMPs, their distribution within the star, and conductive heat transport due to additional scattering events between bound WIMPs and nucleons in the the stellar core. 

Owing to the the negative specific heat of a self-gravitating system, the additional energy means that the cores of dark stars are cooler and less dense than those of normal main sequence stars.  Dark stars are hence larger, cooler and more diffuse than main-sequence stars of similar masses, and in fact resemble protostars lying somewhere on the protostellar cooling (Hayashi) track.  This resemblance can be understood in terms of the differing microscopic properties of the two energy sources.  To a good approximation, the luminosity from WIMP annihilation is decoupled from the gas pressure or temperature, as the capture and annihilation rates are only weakly dependent upon the stellar structure.  The same is true of gravitational potential energy, the primary energy source during protostellar cooling.  Nuclear burning, on the other hand, is strongly dependent upon the core temperature and density.  As WIMP annihilation becomes stronger, nuclear burning becomes progressively less efficient; the more dark matter is present in a stellar core, the more it expands and cools, and the further `back up' the Hayashi track a star can be found in the HR diagram.  This causes an overall decrease in the stellar luminosity for moderate WIMP capture rates, but an increase at large capture rates, where the annihilation energy alone exceeds the equivalent main-sequence luminosity.  The partial replacement of fusion power with energy from dark matter annihilation also slows the rate at which hydrogen is converted to helium, increasing stars' main sequence lifetimes.  This picture is summarised in Fig.~\ref{fig2}, where we give evolutionary tracks for some example main-sequence stars following their immersion in dark matter halos of varying densities.

Whilst the direct physical impacts of dark matter annihilation in stellar cores show only a very weak dependence upon metallicity \cite{Scott09}, the fact that nuclear scattering and gravitational contraction are effective at such different times in a star's life cycle means that their impacts do end up being rather functionally distinct.  In particular, the presence of large amounts of dark matter during the formation of a star (i.e.~as can only be produced due to gravitational contraction) essentially inhibits the collapse, as it can partially prevent further cooling of the gas beyond a certain critical point, well before a main-sequence object has formed \cite{Spolyar08,Natarajan09}.  This means that stars in which gravitational contraction is an effective means of obtaining WIMPs take longer to halt gas accretion by radiative feedback, leading them to acquire larger masses than stars formed without the influence of dark matter.  It is thought that because Pop III stars form near the centre of (almost) pristine dark matter halos, this effect should be essentially unique to Pop III stars.  More speculative work \cite{FreeseSMDS} has even suggested that such a mechanism could lead to the formation of multi-million solar mass `supermassive dark stars'.  In any case, if Pop III dark stars do exist, their larger masses, in combination with the fact that the annihilation luminosity can exceed the equivalent main-sequence luminosity anyway, mean that they are not only cooler and larger than standard Pop III stars of the same mass, but also substantially more luminous.

\section{Impacts at high redshift}

If dark stars exist in the early Universe, they may be observable via their indirect impacts upon cosmological processes, or even directly as extremely red, highly luminous objects.  As single objects, standard Pop III dark stars would only be observable with e,g.~JWST if they were very cool and long-lived, and even then only when viewed through the strongest known gravitational lens \cite{Zackrisson10a}.  Their unique spectral impact on integrated galaxy spectra could however provide a more promising detection channel \cite{Zackrisson10b}.  Supermassive dark stars, on the other hand, should be individually observable by both HST and JWST in many scenarios \cite{Zackrisson10b}, severely constraining their abundance and lifetimes.

\begin{figure}
\centering
\includegraphics[width=0.6\linewidth]{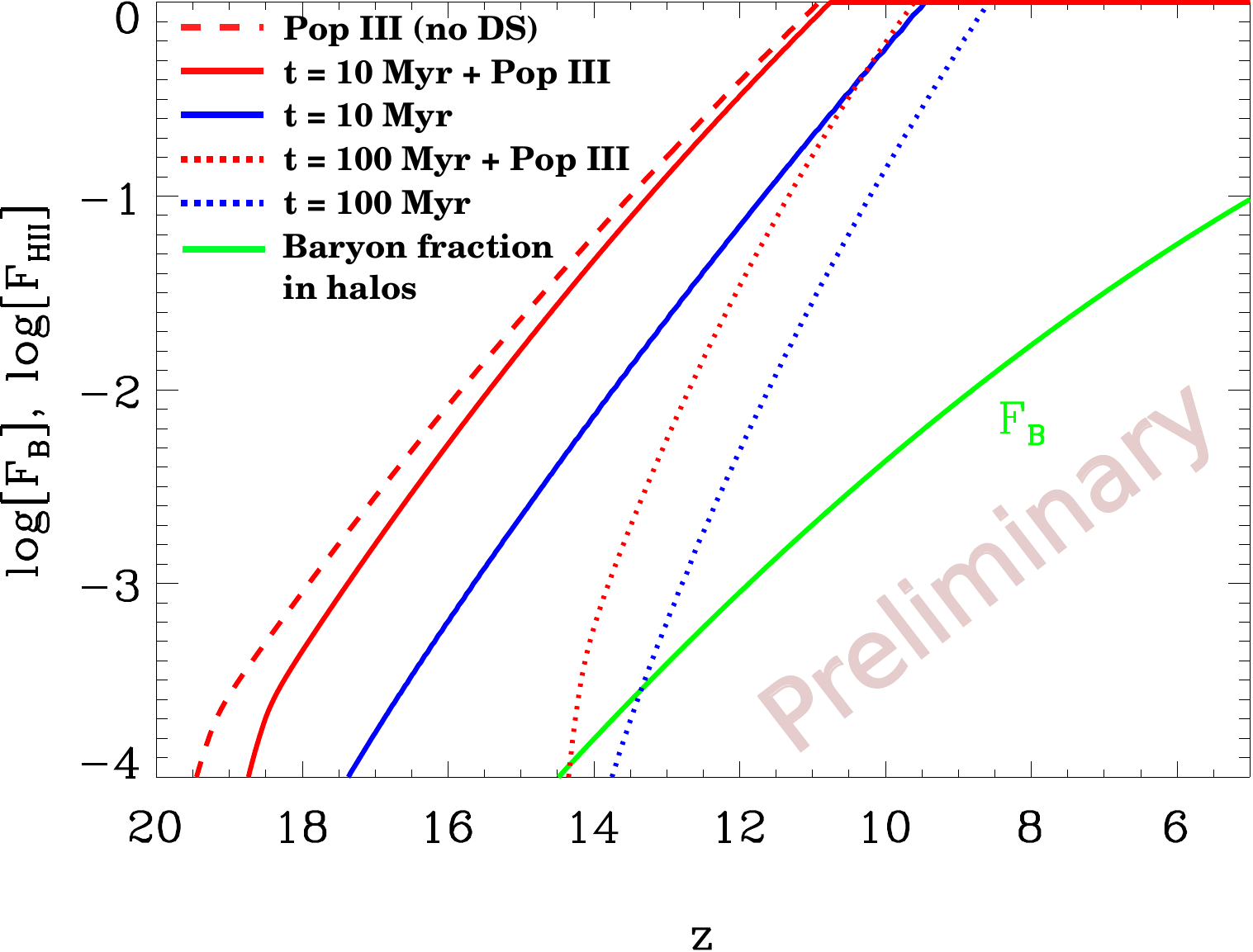}
\caption{Reionisation histories (in terms of the H\,\textsc{ii} fraction) from Ref.~\cite{Venk11}, computed under different assumptions about the populations of dark stars existent in the early Universe.  Dark star curves include an initial population of highly dark-matter-dominated Pop III dark stars, which do not contribute any reionising photons, and hence delay reionisation.  This population is assumed to live for either $t=10$ or $100$\,Myr.  Immediately before their deaths, the dark stars briefly resemble $\sim$800\,$M_\odot$ main-sequence Pop III stars; atmospheres and reionising photon fluxes of dark stars at this stage are modelled using the \textsc{tlusty} code \cite{Hubeny & Lanz}.  After the population of dark stars, reionisation models assume either standard Pop II star formation, or normal Pop III star formation followed by normal Pop II star formation.  The normal Pop III star formation is modelled as having a duration of 10\,Myr, and consisting of stars with masses of $100$--$300\,M_\odot$.  The `no DS' model consists of just this normal Pop III for a duration of 10 Myr, followed by a normal Pop II. The collapsed baryon fraction in halos is also shown for comparison.}
\label{fig3}
\end{figure}

Because they can be more massive (and hence more luminous) than normal Pop III stars, and potentially live for far longer, Pop III dark stars could substantially change the reionisation history of the Universe.  Exactly how reionisation is changed depends not only on the abundance and lifetimes of the dark stars, but to what extent they actually remain affected by WIMP annihilation over the course of the lifetimes.  This is because large scattering-mediated WIMP capture rates lead to very cool dark stars, which do not produce a substantial number of ionising photons for most of their lives.  This causes reionisation to be delayed relative to standard Pop III scenarios, as shown in Fig.~\ref{fig3} \cite{Venk11}.  Here the evolution of the ionised hydrogen gas fraction in the Universe is shown as a function of redshift for a number of different stellar population models, with and without very cool dark stars.  The presence of such cool dark stars can be seen to delay reionisation by different amounts, depending upon the dark star lifetimes and whether they are followed by a `normal' (non-dark) Pop III starburst, or simply by standard Pop II star formation.

On the other hand, in the absence of large scattering-mediated capture rates, the larger stellar masses of dark stars compared to normal Pop III mean that they produce \emph{more} ionising radiation than would otherwise be expected.  This would hence have the opposite effect on reionisation, causing the Universe to become fully ionised at much higher redshifts than in the canonical Pop III scenario \cite{Schleicher}.  Studying the integrated optical depth of the CMB with WMAP and Planck should thus help to further constrain the abundance and lifetimes of the Pop III dark stars \cite{Venk11}.

\section*{Acknowledgements}

I thank Joakim Edsj\"o, Malcolm Fairbairn, Paolo Gondolo, Fabio Iocco, Elena Pierpaoli, Claes-Erik Rydberg, Sofia Sivertsson, Aparna Venkatesan and Erik Zackrisson for fruitful discussions and collaboration on dark stars over the last few years.


\begin{thebibliography}{99}
\bibitem{Jungman96}
Jungman, G., Kamionkowski, M., \& Griest, K. 1996, \physrep, \textbf{267}, 195
\bibitem{Bergstrom00}
Bergstr\"om, L. 2000, \repprogphys, \textbf{63}, 793
\bibitem{Bertone05}
Bertone, G., Hooper, D., \& Silk, J. 2005, \physrep, \textbf{405}, 279
\bibitem{SalatiSilk}
{Salati}, P., \& {Silk}, J. 1989, \apj, \textbf{338}, 24
\bibitem{BouquetSalati}
Bouquet, A., \& Salati, P. 1989, \aap, \textbf{217}, 270
\bibitem{MoskalenkoWai}
Moskalenko, I.~V., \& Wai, L.~L. 2007, \apj, \textbf{659}, L29
\bibitem{Bertone08}
Bertone, G. \& Fairbairn, M. 2008, \prd, \textbf{77}, 043515
\bibitem{McCullough10}
McCullough, M., \& Fairbairn, M. 2010, \prd, \textbf{81}, 083520
\bibitem{Hooper10}
Hooper, D., Spolyar, D., Vallinotto, A., \& Gnedin, N.~Y. 2010, \prd, \textbf{81}, 103531
\bibitem{Fairbairn10}
de Lavallaz, A., \& Fairbairn, M. 2010, \prd, \textbf{81}, 123521
\bibitem{Kouvaris10}
Kouvaris, C., \& Tinyakov, P. 2010, \prd, \textbf{82}, 063531
\bibitem{Scott07}
{Scott}, P., {Edsj{\"o}}, J., \& {Fairbairn}, M. 2008, in \emph{Dark Matter in Astroparticle and Particle Physics: Dark 2007}, ed. H.~K. {Klapdor-Kleingrothaus} \& G.~F. {Lewis} (World Scientific, Singapore), 387 (arXiv:0711.0991)
\bibitem{Fairbairn08}
{Fairbairn}, M., {Scott}, P., \& {Edsj{\"o}}, J. 2008, \prd, \textbf{77}, 047301
\bibitem{Scott09}
{Scott}, P., {Fairbairn}, M., \& {Edsj{\"o}}, J. 2009a, \mnras, \textbf{394}, 82; erratum ibid. 2010, \mnras, \textbf{400}, 2207
\bibitem{Casanellas09}
{Casanellas}, J., \& {Lopes}, I. 2009, \apj, \textbf{705}, 135
\bibitem{Scott09proc}
{Scott}, P., {Edsj{\"o}}, J., \& {Fairbairn}, M. 2009b, in \emph{Dark Matter in Astroparticle and Particle Physics: Dark 2009}, ed. H.~K. {Klapdor-Kleingrothaus} \& I.~V. {Krivosheina} (World Scientific, Singapore), 320, (arXiv:0904.2395)
\bibitem{Spolyar08}
Spolyar, D., Freese, K., \& Gondolo, P. 2008, \prl, \textbf{100}, 051101
\bibitem{Iocco08a}
Iocco, F. 2008, \apj, \textbf{677}, L1
\bibitem{Freese08a}
{Freese}, K., {Spolyar}, D., \& {Aguirre}, A. 2008, JCAP, \textbf{11}, 14
\bibitem{Iocco08b}
Iocco, F., Bressan, A., Ripamonti, E., Schneider, R., Ferrara, A., \& Marigo, P. 2008, MNRAS, \textbf{390}, 1655
\bibitem{Freese08b}
Freese, K, Bodenheimer, P., Spolyar, D., Gondolo, P. 2008, \apj, \textbf{685}, L101
\bibitem{Yoon08}
Yoon, S.-C., Iocco, F., Akiyama, S. 2008, \apj, \textbf{688}, L1
\bibitem{Taoso08}
Taoso, M., Bertone, G., Meynet, G., \& Ekstr\"om, S. 2008, \prd, \textbf{78}, l3510
\bibitem{Freese09}
Freese, K., Gondolo, P., Sellwood, J. A., \& Spolyar, D. 2009, \apj, \textbf{693}, 1563
\bibitem{Natarajan09}
Natarajan, A., Tan, J. C., \& O'Shea, B. W. 2009, \apj, \textbf{692}, 574
\bibitem{Ripamonti09}
Ripamonti, E., Iocco, F., Bressan, A., Schneider, R., Ferrara, A, \& Marigo, P. 2009, in \emph{Identification of Dark Matter 2008}, PoS(idm2008)073 
\bibitem{Spolyar09}
Spolyar, D., Bodenheimer, P., Freese, K., \& Gondolo, P. 2009, \apj, \textbf{705}, 1031 
\bibitem{Iocco09}
Iocco, F. 2009, Nucl.\ Phys.\ Proc.\ Suppl., \textbf{194}, 82
\bibitem{Umeda10}
{Umeda}, H., {Yoshida}, N., {Nomoto}, K., {Tsuruta}, S., {Sasaki}, M., \& {Ohkubo}, T. 2009, \jcap, \textbf{8}, 24
\bibitem{Sivertsson10}
Sivertsson, S., \& Gondolo, P. 2010, \prd, in press, (arXiv:1006.0025)
\bibitem{Ripamonti10}
Ripamonti, E., Iocco, F., Bressan, A., Schneider, R., Ferrara, A, \& Marigo, P. 2010, \mnras, \textbf{406}, 2605 
\bibitem{Gondolo10}
Gondolo, P., Huh, J-H., Do Kim, H., \& Scopel, S. \jcap, \textbf{7}, 26
\bibitem{Schleicher}
Schleicher, D. R. G., Banerjee, R., \& Klessen, R. S. 2009, \prd, \textbf{79}, 3510
\bibitem{Venk11}
Venkatesan, A., Scott, P., Gondolo, P., \& Pierpaoli, E., in preparation
\bibitem{Zackrisson10a}
Zackrisson, E., Scott, P., Rydberg, C-E., Iocco, F., Edvardsson, B., \"Ostlin, G., Sivertsson, S., Zitrin, A., Broadhurst, T., Gondolo, P. 2010, \apj, \textbf{717}, 257
\bibitem{FreeseSMDS}
Freese, K., Ilie, C., Spolyar, D., Valluri, M., \& Bodenheimer, P. 2010, \textbf{716}, 1397
\bibitem{Zackrisson10b}
Zackrisson, E., Scott, P., Rydberg, C-E., Iocco, F., Sivertsson, S., \"Ostlin, G., Mellema, G., Iliev, I.~T., Shapiro, P.~R. 2010, \mnras, \textbf{407}, L74
\bibitem{Hubeny & Lanz}
Hubeny, I., \& Lanz, T. 1995, \apj, \textbf{439}, 875
\end{thebibliography}
\end{document}